\documentclass{llncs}
\usepackage{makeidx}  
\usepackage{amssymb}
\usepackage{amsmath}
\usepackage{url}
\usepackage{pstricks, pst-node, pst-tree}

\usepackage{lastpage}

\usepackage{fancyhdr} 
\makeatletter
\renewcommand{\@evenfoot}%
{\hfil{\thepage} of \pageref{LastPage}\hfil}
\renewcommand{\@oddfoot}{\@evenfoot}
\makeatother

\begin{document}
\mainmatter              
\title{Computing Fibonacci numbers on a Turing Machine}
\author{Alex Vinokur}
%
%
\tocauthor{Holon, Israel}
\institute{Holon, Israel\\
\email{alexvn@barak-online.net}\\
\email{alex.vinokur@gmail.com}\\ Home Page:
\texttt{http://alexvn.freeservers.com/}}

\maketitle              

\begin{abstract}
A Turing machine that computes Fibonacci numbers is described.
\end{abstract}
\section{Preface}

The program computes a Fibonacci number.

\

A number $n$ is represented by $n$ 1-s.

{\bfseries{Sample}} :

5 is represented as 1 1 1 1 1

3 is represented as 1 1 1

\

\underline{Input} : number $n$

{\bfseries{Sample}} (n = 7) :

1 1 1 1 1 1 1

\

\underline{Output} : Fibonacci-$n$

{\bfseries{Sample}} (Fibonacci-7) :

1 1 1 1 1 1 1 1 1 1 1 1 1

\newpage
\section{Alphabets of States and Symbols}

Here are alphabets of states and symbols.

\
\

\underline{State alphabet}

Initial state: $q_0$;

Halting state: $q_f$;

Internal states: 
$q_{101}$, $q_{102}$, $q_{103}$, $q_{104}$, $q_{105}$, $q_{106}$, $q_{107}$, $q_{108}$, $q_{109}$, 
$q_{201}$, $q_{202}$, $q_{203}$, $q_{204}$, 
$q_{301}$, $q_{302}$, $q_{303}$, $q_{304}$, $q_{305}$, $q_{306}$, $q_{307}$, $q_{308}$, $q_{309}$, $q_{310}$, $q_{311}$, 
$q_{401}$, $q_{402}$, $q_{403}$, $q_{404}$, 
$q_{501}$, $q_{502}$, $q_{503}$, 
$q_{601}$, $q_{602}$, $q_{603}$, $q_{604}$, 
$q_{701}$, $q_{702}$, $q_{703}$, $q_{704}$, 
$q_{801}$, $q_{802}$, $q_{803}$, $q_{804}$, $q_{805}$, $q_{806}$, $q_{807}$, $q_{808}$, $q_{809}$.

\
\

\underline{Symbol alphabet}

Empty symbols alphabet: $b$;

Input alphabet: $1$;

Internal alphabet: $x$, $*$.

\newpage
\section{Transition Table}

The table contains 100 rules.

\begin{center}

\begin{tabular}{ c | c c | c c c }
\hline
$Rule_{No}$&$State_{cur}$&$Symbol_{cur}$&$State_{next}$&$Symbol_{next}$&$Head_{move}$\\
\hline
 0 & $q_0$     & $1$ & $q_{101}$ & $x$ & $R$\\
 1 & $q_{101}$ & $1$ & $q_{101}$ & $1$ & $R$\\
 2 & $q_{101}$ & $b$ & $q_{102}$ & $1$ & $R$\\
 3 & $q_{102}$ & $b$ & $q_{103}$ & $*$ & $R$\\
 4 & $q_{103}$ & $b$ & $q_{104}$ & $1$ & $R$\\
 5 & $q_{104}$ & $b$ & $q_{601}$ & $*$ & $L$\\
 6 & $q_{105}$ & $b$ & $q_{106}$ & $1$ & $L$\\
 7 & $q_{106}$ & $*$ & $q_{701}$ & $*$ & $L$\\
 8 & $q_{107}$ & $*$ & $q_{108}$ & $*$ & $L$\\
 9 & $q_{107}$ & $1$ & $q_{107}$ & $1$ & $L$\\
10 & $q_{108}$ & $*$ & $q_{109}$ & $*$ & $N$\\
11 & $q_{108}$ & $1$ & $q_{108}$ & $1$ & $L$\\
12 & $q_{109}$ & $*$ & $q_{109}$ & $*$ & $R$\\
13 & $q_{109}$ & $1$ & $q_{109}$ & $1$ & $R$\\
14 & $q_{109}$ & $b$ & $q_{201}$ & $*$ & $N$\\
15 & $q_{201}$ & $*$ & $q_{202}$ & $*$ & $L$\\
16 & $q_{201}$ & $1$ & $q_{201}$ & $1$ & $L$\\
17 & $q_{202}$ & $*$ & $q_{203}$ & $*$ & $R$\\
18 & $q_{202}$ & $1$ & $q_{202}$ & $1$ & $L$\\
19 & $q_{202}$ & $b$ & $q_{203}$ & $b$ & $R$\\
20 & $q_{203}$ & $*$ & $q_{301}$ & $*$ & $N$\\
21 & $q_{203}$ & $1$ & $q_{204}$ & $b$ & $R$\\
22 & $q_{204}$ & $*$ & $q_{204}$ & $*$ & $R$\\
23 & $q_{204}$ & $1$ & $q_{204}$ & $1$ & $R$\\
24 & $q_{204}$ & $b$ & $q_{201}$ & $1$ & $L$\\
25 & $q_{301}$ & $*$ & $q_{302}$ & $*$ & $L$\\
26 & $q_{302}$ & $*$ & $q_{303}$ & $*$ & $L$\\
27 & $q_{302}$ & $b$ & $q_{302}$ & $b$ & $L$\\
28 & $q_{303}$ & $*$ & $q_{304}$ & $*$ & $R$\\
29 & $q_{303}$ & $1$ & $q_{303}$ & $1$ & $L$\\
30 & $q_{303}$ & $b$ & $q_{304}$ & $b$ & $R$\\
31 & $q_{304}$ & $*$ & $q_{308}$ & $b$ & $N$\\
32 & $q_{304}$ & $1$ & $q_{305}$ & $b$ & $R$\\
33 & $q_{305}$ & $*$ & $q_{306}$ & $*$ & $R$\\
34 & $q_{305}$ & $1$ & $q_{305}$ & $1$ & $R$\\
35 & $q_{306}$ & $*$ & $q_{307}$ & $*$ & $L$\\
36 & $q_{306}$ & $1$ & $q_{307}$ & $1$ & $L$\\
37 & $q_{306}$ & $b$ & $q_{306}$ & $b$ & $R$\\
38 & $q_{307}$ & $b$ & $q_{302}$ & $1$ & $L$\\
39 & $q_{308}$ & $1$ & $q_{309}$ & $1$ & $L$\\
\hline

\end{tabular}

\

Table continued on next page

\newpage

Continued

\

\begin{tabular}{ c | c c | c c c }
\hline
$Rule_{No}$&$State_{cur}$&$Symbol_{cur}$&$State_{next}$&$Symbol_{next}$&$Head_{move}$\\
\hline

40 & $q_{308}$ & $b$ & $q_{308}$ & $b$ & $R$\\
41 & $q_{309}$ & $b$ & $q_{310}$ & $*$ & $L$\\
42 & $q_{310}$ & $*$ & $q_{311}$ & $*$ & $R$\\
43 & $q_{310}$ & $b$ & $q_{310}$ & $1$ & $L$\\
44 & $q_{311}$ & $*$ & $q_{501}$ & $*$ & $R$\\
45 & $q_{311}$ & $1$ & $q_{311}$ & $1$ & $R$\\
46 & $q_{401}$ & $*$ & $q_{402}$ & $*$ & $L$\\
47 & $q_{401}$ & $1$ & $q_{401}$ & $1$ & $L$\\
48 & $q_{402}$ & $*$ & $q_{403}$ & $*$ & $L$\\
49 & $q_{402}$ & $1$ & $q_{402}$ & $1$ & $L$\\
50 & $q_{403}$ & $*$ & $q_{403}$ & $*$ & $L$\\
51 & $q_{403}$ & $1$ & $q_{404}$ & $*$ & $L$\\
52 & $q_{404}$ & $*$ & $q_{404}$ & $*$ & $R$\\
53 & $q_{404}$ & $1$ & $q_{404}$ & $1$ & $R$\\
54 & $q_{404}$ & $b$ & $q_{201}$ & $*$ & $N$\\
55 & $q_{404}$ & $x$ & $q_{801}$ & $x$ & $N$\\
56 & $q_{501}$ & $*$ & $q_{502}$ & $1$ & $N$\\
57 & $q_{501}$ & $1$ & $q_{501}$ & $1$ & $R$\\
58 & $q_{502}$ & $1$ & $q_{502}$ & $1$ & $R$\\
59 & $q_{502}$ & $b$ & $q_{503}$ & $b$ & $L$\\
60 & $q_{503}$ & $1$ & $q_{401}$ & $b$ & $L$\\
61 & $q_{601}$ & $*$ & $q_{602}$ & $*$ & $L$\\
62 & $q_{601}$ & $1$ & $q_{601}$ & $1$ & $L$\\
63 & $q_{602}$ & $1$ & $q_{603}$ & $*$ & $L$\\
64 & $q_{603}$ & $1$ & $q_{604}$ & $1$ & $R$\\
65 & $q_{603}$ & $x$ & $q_{801}$ & $x$ & $N$\\
66 & $q_{604}$ & $*$ & $q_{604}$ & $*$ & $R$\\
67 & $q_{604}$ & $1$ & $q_{604}$ & $1$ & $R$\\
68 & $q_{604}$ & $b$ & $q_{105}$ & $b$ & $N$\\
69 & $q_{701}$ & $*$ & $q_{702}$ & $*$ & $L$\\
70 & $q_{701}$ & $1$ & $q_{701}$ & $1$ & $L$\\
71 & $q_{702}$ & $*$ & $q_{702}$ & $*$ & $L$\\
72 & $q_{702}$ & $1$ & $q_{703}$ & $*$ & $L$\\
73 & $q_{703}$ & $1$ & $q_{704}$ & $1$ & $R$\\
74 & $q_{703}$ & $x$ & $q_{801}$ & $x$ & $N$\\
75 & $q_{704}$ & $*$ & $q_{704}$ & $*$ & $R$\\
76 & $q_{704}$ & $1$ & $q_{704}$ & $1$ & $R$\\
77 & $q_{704}$ & $b$ & $q_{107}$ & $b$ & $L$\\
78 & $q_{801}$ & $*$ & $q_{801}$ & $*$ & $R$\\
79 & $q_{801}$ & $1$ & $q_{801}$ & $1$ & $R$\\
\hline

\end{tabular}

\

Table continued on next page

\newpage

Continued

\

\begin{tabular}{ c | c c | c c c }
\hline
$Rule_{No}$&$State_{cur}$&$Symbol_{cur}$&$State_{next}$&$Symbol_{next}$&$Head_{move}$\\
\hline

80 & $q_{801}$ & $b$ & $q_{802}$ & $b$ & $L$\\
81 & $q_{801}$ & $x$ & $q_{801}$ & $x$ & $R$\\
82 & $q_{802}$ & $*$ & $q_{808}$ & $b$ & $L$\\
83 & $q_{802}$ & $1$ & $q_{808}$ & $1$ & $L$\\
84 & $q_{803}$ & $*$ & $q_{803}$ & $*$ & $L$\\
85 & $q_{803}$ & $1$ & $q_{803}$ & $*$ & $L$\\
86 & $q_{803}$ & $x$ & $q_{804}$ & $x$ & $R$\\
87 & $q_{804}$ & $*$ & $q_{804}$ & $*$ & $R$\\
88 & $q_{804}$ & $1$ & $q_{805}$ & $*$ & $L$\\
89 & $q_{804}$ & $b$ & $q_{809}$ & $b$ & $N$\\
90 & $q_{805}$ & $*$ & $q_{805}$ & $*$ & $L$\\
91 & $q_{805}$ & $1$ & $q_{806}$ & $1$ & $R$\\
92 & $q_{805}$ & $x$ & $q_{806}$ & $*$ & $N$\\
93 & $q_{806}$ & $*$ & $q_{807}$ & $1$ & $R$\\
94 & $q_{807}$ & $*$ & $q_{804}$ & $*$ & $R$\\
95 & $q_{808}$ & $*$ & $q_{803}$ & $*$ & $L$\\
96 & $q_{808}$ & $1$ & $q_{808}$ & $1$ & $L$\\
97 & $q_{809}$ & $*$ & $q_{809}$ & $b$ & $L$\\
98 & $q_{809}$ & $1$ & $q_{f}  $ & $1$ & $N$\\
99 & $q_{809}$ & $b$ & $q_{809}$ & $b$ & $L$\\
\hline

\end{tabular}

\end{center}

\section{Testing the Machine}
C++ Simulator of a Turing machine has been used to compute several Fibonacci numbers.
The simulator can be downloaded at
\begin{itemize}
\item  {\url{http://sourceforge.net/projects/turing-machine}}
\item  {\url{http://alexvn.freeservers.com/s1/turing.html}}
\end{itemize}

Raw logs can be seen at \url{http://groups.google.com/groups?th=1e653c4ef60faa44}

\end{document}